\documentclass[aps,prb,twocolumn,amsmath,amssymb,floatfix, showpacs]{revtex4-1}
 
 \usepackage{graphicx}
\usepackage{dcolumn} 
\usepackage{bm}      
\usepackage[usenames,dvipsnames]{color}
\usepackage{ulem} 

\begin{document}
\title{Propagating and annihilating vortex dipoles in the Gross-Pitaevskii equation}  
\author{Cecilia Rorai$^{1,2,3}$,  K. R. Sreenivasan$^{2,4}$ and Michael E. Fisher$^{2}$}
\affiliation{ $^{1}$Institute for Research in Electronics and Applied Physics,}
\affiliation{$^{2}$Institute for Physical Science and Technology, University of Maryland, College Park, Maryland 20740, USA}
\affiliation{$^{3}$Doctorate School in Environmental and Industrial Fluid Mechanics, Universit\'{a} degli Studi di Trieste, Trieste, Italy}
\affiliation{$^{4}$Department of Physics and Courant Institute of Mathematical Sciences, New York University, New York 10012, USA.}
 
\begin{abstract}
Quantum vortex dynamics in Bose-Einstein condensates or superfluid helium can be informatively described by the Gross-Pitaevskii (GP) equation.
Various approximate analytical formulae for a single stationary vortex are recalled and their shortcomings demonstrated. Significantly more accurate two-point [2/2] and [3/3] Pad\'{e} approximants for stationary vortex profiles are presented. Two straight, singly quantized, antiparallel vortices, located  at a distance  $d_0$  apart, form a vortex dipole, which, in the GP model, can {\it either} annihilate {\it or} propagate indefinitely as a `solitary wave'.
We show, through calculations performed in a periodic domain, that the details and types of behavior displayed by vortex dipoles depend strongly on the initial conditions rather than only on the separation distance $d_0$ (as has been previously claimed).
It is found, indeed, that the choice of the initial two-vortex profile (i.e., the modulus of the `effective wave function'), strongly affects the vortex trajectories and the time scale of the process: annihilation proceeds more rapidly when low-energy (or `relaxed') initial profiles are imposed. The initial `circular' {\it phase distribution contours}, customarily obtained by multiplying an effective wave function for each individual vortex, can be generalized to explicit {\it elliptical} forms specified by two parameters; then  by `tuning' the elliptical shape at fixed $d_0$, a sharp transition between solitary-wave propagation and annihilation is captured. Thereby, a `phase diagram' for this `AnSol' transition is constructed in the space of ellipticity and separation and various limiting forms of the boundary are discussed.

\end{abstract}
\maketitle

\section{Introduction}
The Gross-Pitaevskii (GP) equation \cite{ginzburg,  pitaevskii, Gross} is widely accepted as a basic model to study vortex dynamics in superfluid helium or Bose Einstein condensates. It has the merit of providing an effective quantum mechanical description of the vortex core dynamics, while also allowing detailed predictions of vortex reconnection,\cite{koplik93, koplik96, leadbeater01, Tsubota02, Tebbs, Bajer11, Kerr_PRL} a phenomenon which is thought to centrally affect the large scale behavior of quantum turbulence.\cite{paoletti, White} 

The general focus of this present work is the specification of appropriate initial conditions to perform systematic, reproducible, vortex dynamics calculations through the GP equation. This is a matter of some delicacy, because given the nonlinearities involved one must rely almost entirely on numerical solutions, which, as we will demonstrate, depend strongly on the details of the initially imposed complex, effective wavefunction $\Psi({\bf x}, t=0)$. Of course, the boundary conditions also play a role, although for the present purposes, a less crucial one. In this study we will always assume periodic boundary conditions.

We address the general objective here by studying single vortex profiles and {\it vortex dipole} dynamics, which may be usefully relevant to physical situations that can be regarded as two-dimensional,  such as helium films,\cite{siggia80} or experiments performed in trapped Bose-Einstein condensates.\cite{schumayer07, neely10, freilich10, aioi11}

In order to select initial conditions that lead to simulated vortex dynamics, two conceptual steps are needed: first, one needs to describe mathematically a quantum vortex appropriately for the GP equation; second, one needs to set up one or more vortices in the computational domain. 

This second step has usually been achieved merely by multiplying together wave functions describing each individual vortex.\cite{koplik93, koplik96,Tsubota02, Yepez} For exploratory purposes this device is simple and may well be adequate. However, if one wishes to address more quantitative and subtle issues, such as arise naturally in contemplating the wealth of experimental data \cite{bewley, paoletti, paoletti09} and what may be found in the future, a more systematic approach is called for.  Here we describe a different attack which exploits the properties of the diffusion equation associated with the GP equation and thus allows one to impose {\it low-energy initial conditions}. The use of the diffusive GP equation is sometimes referred to as the {\it imaginary time propagation method}.\cite{Dalfovo, Lehtovaara, White} 

The mathematical description even of a {\it single vortex} entails some complications. A single straight vortex in an infinite domain is, in fact, optimally represented as a solution of the stationary GP equation. However, there is no convenient analytical form to express it precisely; as a result, a numerical study is required. However, this may not be convenient or practical when wave functions for multiple vortices are required. Thus it has been customary to employ some analytically convenient but intrinsically approximate profiles. Various approximate profiles for a single straight vortex have been proposed. We recall three of these, namely, in order of increasing complexity, the Fetter,\cite{fetter} Kerr \cite{Kerr_PRL} and Berloff \cite{berloff2} approximants; however, as distinct improvements we propose new [2/2] and [3/3] Pad\'{e} approximants. 
It is worth stressing, in this regard, that the GP equation conserves energy. Consequently any energy imposed by means of vortex initial conditions that exceeds the minimal energy entailed in the exact vortex configuration is not dissipated but, rather, dispersed by mechanisms which are not easy to predict {\it a priori}, and often confusing to interpret.

In the GP equation unit {\it vortex dipoles}, characterized by a separation distance $d_{0}$, can either annihilate and emit their energy via outgoing waves, or, after a transient, propagate at constant velocity in the form of solitary waves. Furthermore, Ivonin \cite{ivonin97} and Ogawa {\it et al.}\cite{Tsubota02} claimed there was a nonzero critical value of  $d_{0}$,  below which no solutions describing dipoles moving uniformly were possible. However, Jones and Roberts\cite{jones81} and Berloff  \cite{berloff2, berloff04} found accurate numerical solutions for constant velocity solitary waves at separations down to the limit\cite{berloff2} $d_{0}=0$, which corresponded to a propagation velocity $U=0.45$, in dimensionless units where the speed of sound is\cite{ivonin97, Tsubota02} $c=1/\sqrt{2}$. As we show, this discrepancy highlights the importance of the initial vortex configurations in studying solutions of the GP equation. In particular, the phase pattern of the initial order parameter is significant. Indeed, by varying it, we are able to generate solutions in which the vortex dipoles either annihilate {\it or} propagate. Our GP calculations also serve to check the dependence of the time-to-annihilation and of the dipole propagation velocity, $U$, on the initial separation as $d_{0}$ $\rightarrow$ 0.

In Sec. II we recall the Gross-Pitaevskii and we describe the Diffusive Gross-Pitaevskii (DGP) equation. Section III is devoted to the study of single vortex profiles, while in Sec. IV we present some illustrative vortex dipole calculations that demonstrate how different choices of even very similar initial conditions significantly affect aspects of the observed phenomenon. 
Finally our results are summarized in Sec. V.

\section{The Gross-Pitaevskii equation}
The Gross-Pitaevskii equation,\cite{ginzburg, pitaevskii, Gross} expressed in terms of an effective complex bosonic or `condensate' wave function $\Psi({\bf x},t)$, where ${\bf x}$ is a $d$(=2 or 3)-dimensional spatial coordinate with $t$ denoting the time, is
\begin{equation}
i\hbar \frac{\partial \Psi}{\partial t}= - \frac{\hbar^{2}}{2m}\nabla^{2}\Psi+V_{0}\Psi |\Psi|^{2} -\mu \Psi.\label{GP}
\end{equation}
Here $2\pi\hbar$ is Planck's constant, $m$ is the mass of the Bose particles being simulated, while $\mu$ is the chemical potential and $V_{0}$ represents the strength of the short-range boson-boson repulsive potential. Evidently, the nonlinear term accounts for the interatomic interactions. 

This equation can be made dimensionless by rescaling the coordinates $\bf x$, $ t$, and $\Psi({\bf x},t)$, respectively, by a characteristic length, the healing length\cite{Gross} $\xi_0=\hbar/\sqrt{2m\mu}$ (of order $0.5$  $\AA $ for helium-4), by a characteristic time,  $t_{0}=m\xi_0^{2}/\hbar$ (of order $10^{-1} $ ps for helium-4), and by a characteristic modulus, chosen as the value of the wave function in an infinite domain prior to the onset of a disturbance\cite{Berloff} $|\Psi_{\infty}|=\sqrt{\mu/V_{0}}$. The dimensionless equation then becomes
\begin{equation}
-2i\frac{\partial \Psi}{\partial t}=\nabla^{2}\Psi+(1-|\Psi|^{2})\Psi.  \label{nondim}
\end{equation}
The total energy, $E$, associated with this form, measured inside a domain $\Omega$ with respect to the uniform state $\Psi_{\infty}=1$ is conserved, and is given by the sum of the kinetic energy $E_{K}$, and the potential energy $E_{I}$
with\cite{Berloff}
\begin{equation}
E_{K}=\frac{1}{2}\int_{\Omega} |\nabla \Psi|^{2} d{\bf x}  
\end{equation}
and
\begin{equation}
E_{I}=\frac{1}{4}\int_{\Omega}(1-|\Psi|^{2})^{2}d{\bf x}. \label{energy}
\end{equation}

The kinetic energy can be further split into a classical kinetic energy, $E_{CK}$, and a quantum energy, $E_{Q}$, which are given by, respectively,\cite{Berloff}

\begin{equation}
E_{CK}=\frac{1}{2}\int_{\Omega} f^2(\nabla \phi)^2 d{\bf x}  
\end{equation}
where $\Psi$ is expressed as $\Psi=f({\bf x}, t)e^{i\phi({\bf x}, t)}$, and\begin{equation}
E_{Q}=\frac{1}{2}\int_{\Omega}(\nabla f)^2 d{\bf x}. \label{energy}
\end{equation}
It is useful, for thinking in physical terms, to define the density of the condensate $\rho$ as $\rho=f^2$, and the velocity field ${\bf v}$ as ${\bf v}=\nabla \phi$. These equivalences allow to write the GP equation in a hydrodynamical form.\cite{donnelly} 

The diffusion equation associated with the GP equation -- the DGP equation -- is 
\begin{equation}
2\frac{\partial \Psi}{\partial t}=\nabla^{2}\Psi+(1-|\Psi|^{2})\Psi.  \label{diff}
\end{equation}
This does not conserve energy; rather, when $t \rightarrow \infty$, one has $\Psi \rightarrow \bar{\Psi}$, where $\bar{\Psi}$ minimizes the total energy and coincides with a {\it stationary solution} of the GP equation. Therefore, the DGP equation can be used to find fixed points for the GP equation.\cite{Meichle} (As noted above, this imaginary time propagation method, has been often used by the BEC community.)\cite{Dalfovo, Lehtovaara} 

However, even if the focus is rather on the dynamical behavior of unstable structures, such as vortex dipoles or antiparallel vortices with three-dimensional geometries, the DGP equation can be conveniently exploited to find low-energy initial conditions. Here this is demonstrated for vortex dipoles. 

In our calculations,\cite{mythesis} the GP equation has been integrated numerically through a split-step spectral method \cite{Yang} in a periodic domain. The use of the Discrete Cosine Transform (DCT), rather than the Discrete Fourier Transform (DFT), allows one to account for periodic images by calculations in a domain smallerÊ by a factor of four (in two dimensions). The DCT, indeed, implies an even extension of the original function which saves us from having to add three reflections of each single vortex in the computational box, a condition otherwise essential to meet the appropriate periodicity conditions.\cite{mythesis} The initial conditions (before any relaxation with the DGP equation) are obtained by multiplying the wave-function that represents the vortices inside the domain by the wave-functions that correspond to the first eight periodic images outside the domain to ensure that $\Psi$ at the boundaries is sufficiently smooth.\cite{koplik93, Kerr_PRL, mythesis} 

\section{Vortex profiles} 
\begin{figure*}
\resizebox{16.0cm}{!}{\includegraphics{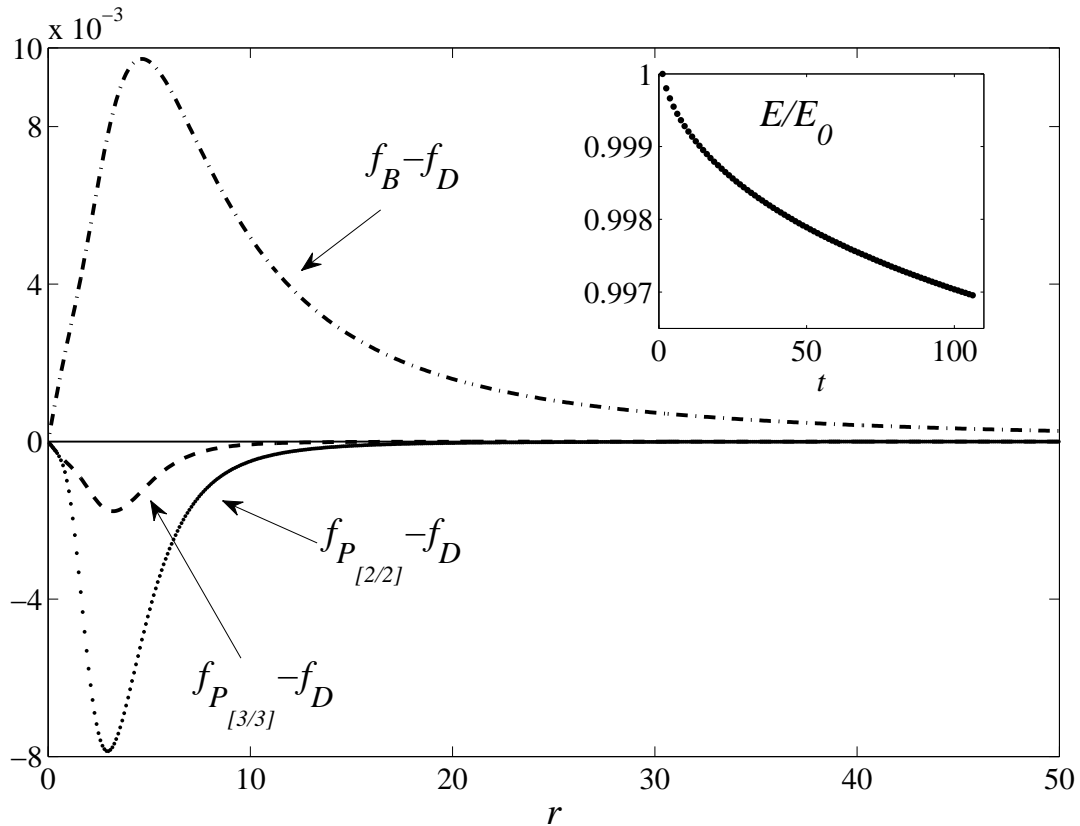}}
\caption{\label{pade_comparison}  Difference between the Berloff approximant and the diffused solution (dash-dotted line), between the [2/2] Pad\'{e} approximant and the diffused profile (dotted line), and between the [3/3] Pad\'{e} approximant and the diffused profile (dashed line). Inset: time decay of the ratio between the total energy $E$ and the total initial energy $E_{0}$ for the [2/2] Pad\'{e} approximant evolved through the DGP equation.}
\end{figure*}
\begin{table*}
  \centering 
  \begin{tabular}{|c|c|}
\hline
{\bf Profile} & {$\bf L_{0}$}\\
\hline
$f_{F}$   & 0.403\\
\hline
 $f_{K}$  & 0.604\\
 \hline
$f_{B}$   & 0.390\\
\hline
$f_{P[2/2]}$ & 0.382\\
\hline
$f_{P[3/3]}$ & 0.381\\
\hline
$f_{D}$   & 0.377\\
\hline
\end{tabular}
  \caption{$L_{0}$ parameter for the five different approximate profiles and the numerical diffused profile.}\label{L0}
\end{table*}

In the language of the GP equation, a vortex line is a long, thin, locally cylindrical structure, straight, bent or wavy, defined by a linear locus of zeros of the effective wavefunction. In particular, a quiescent straight vortex is represented by a stationary axisymmetric solution of (\ref{nondim}) which can be written in cylindrical coordinates, ($r$, $\phi$, $z$), as
\begin{equation}
\Psi(r, \phi, z)=f(r)e^{\pm i \phi n},  \label{vortex_adim}
\end{equation}
where we may term $f(r)$ the `profile' while the phase $\phi$ coincides with the azimuthal angle and $n$ is an integer. 
Using the GP equation (\ref{nondim}) yields
\begin{equation}
f=-\left[\frac{1}{r}\frac{d}{dr}\left(r\frac{d}{dr}\right)-\frac{n^{2}}{r^{2}}\right]f+ f |f|^{2}, \label{radial}
\end{equation}
which can be integrated numerically in an infinite domain\cite{ginzburg, pitaevskii} to determine the nature of $f(r)$. The boundary conditions require that $f(r) \rightarrow 1$ as $r \rightarrow \infty$, while, in the core of the vortex, $f(r) \rightarrow 0$ as $r \rightarrow 0$. 

Since, as mentioned, there is no analytical expression for $f(r)$, various approximate formulae have been proposed. Among these are ({\it i}) a [1/1] Pad\'e approximant advanced by Fetter,\cite{fetter} namely, 
\begin{equation}
f_F(r)=\frac{\sqrt{r^{p}}}{\sqrt{r^{p}+2}} \label{pade_approx}
\end{equation}
with $p=2$;
({\it ii}) a `square core' approximant, introduced by Kerr,\cite{Kerr_PRL} say $f_{K}$, also given by (\ref{pade_approx}) but with $p=4$.
This particular choice was introduced as a stratagem to ensure that the initial density went smoothly from zero, on the vortex core, to roughly the background density over the distance of the healing length. \cite{Kerr_PRL} 
Finally, ({\it iii}) Berloff\cite{berloff2} derived a [2/2] Pad\'e approximant specified by
\begin{equation}
f_{B}(r)\simeq\sqrt{\frac{r^{2}(0.3437+0.0286 r^{2})}{1+0.3333r^{2}+0.0286r^{4}}}. \label{berloff_approx}
\end{equation}

A precise stationary solution, which for convenience we will call the diffused profile $f_{D}(r)$, can be found by evolving in time through the DGP equation any one of these three profiles. Since we are imposing periodic boundary conditions, we need a large enough domain to minimize the influence of the images. We have calculated $f_{D}(r)$ in a domain of size $L_{x}=L_{y}=200$ using a grid spacing $\Delta x=\Delta y\simeq 0.098$.

To improve on the options we have generated a two-point [2/2] Pad\'{e} approximant \cite{Baker} in the form 
\begin{equation}
f_{[2/2]}^{2}(r)=\frac{a_{0}+a_{1}r^{2}+a_{2}r^{4}}{1+b_{1}r^{2}+b_{2}r^{4}},\label{new_pade1}
\end{equation}
and a corresponding [3/3] Pad\'{e} approximant
\begin{equation}
f_{[3/3]}^{2}(r)=\frac{c_{0}+c_{1}r^{2}+c_{2}r^{4}+c_{3}r^{6}}{1+d_{1}r^{2}+d_{2}r^{4}+d_{3}r^{6}}.\label{new_pade2}
\end{equation}
By recalling that, at small $r$, the profile $f(r)$ can be expanded as $f(r)\simeq\sum_{i=1}^{\infty}p_{i}r^{2i-1}$, with\cite{berloff2} $p_{1}\simeq0.5827811878$, $p_{2}=-p_{1}/8$, and $p_{3}=p_{1}(p_{1}^{2}+1/8)/24$, while about $r \rightarrow \infty$ one has\cite{berloff2} $f(r)\simeq \sum_{i=0}^{\infty}q_{i}r^{-2i}$, with $q_{0}=1$, $q_{1}=-1/2$ $q_{2}=-9/8$ and $q_{3}=-161/16$, we find, for $f_{[2/2]}$, 
$a_{0}=0$, $a_{1}=p_{1}^{2}$, $a_{2}=b_{2}$, and
\begin{eqnarray}
b_{1}&=&\frac{3p_{1}^2}{4(1-p_{1}^{2})}\simeq 0.3857 \nonumber, \\
 b_{2}&=&p_{1}^2\left[\frac{4p_{1}^2-1}{4(1-p_{1}^2)}\right]\simeq 0.0461.
 \end{eqnarray} 
The expressions for $b_{1}$ and $b_{2}$ are found by expanding (\ref{new_pade1}) in a Taylor series for small $r$ and large $r$ and equating the coefficients of the second term, respectively, to $p_{2}$ and $q_{1}$. This approximant then reproduces $p_{1}$, $p_{2}$ {\it and} $q_{0}$, $q_{1}$ correctly. 

Following an analogous procedure we find, for $f_{[3/3]}$,  $c_{0}=0$, $c_{1}=p_{1}^{2}$, $c_{3}=d_{3}$, $c_{2}\simeq0.0501$, $d_{1}\simeq0.3976$, $d_{2}\simeq0.0527$ and $d_{3}\simeq0.0026$. This approximant reproduces $p_{1}$, $p_{2}$, $p_{3}$ and $q_{0}$, $q_{1}$, $q_{2}$.
The $f_{[2/2]}$ and $f_{[3/3]}$ Pad\'{e} approximants  and the Berloff approximant are compared in Fig. \ref{pade_comparison} with respect to the diffused profile $f_{D}$. The [3/3] Pad\'{e} approximant is the closest to the diffused profile followed by the  [2/2] Pad\'{e} approximant and, lastly, the Berloff approximant.

It is observed that expanding $f_{F}$, $f_{K}$ and $f_{B}$ in power series, and examining the limits for small and large $r$, yields a variety of results, none of which reproduces $p_{1}$, $p_{2}$ and $q_{1}$. Explicitly, for small $r$, one has
\begin{eqnarray}
f_{F}&\simeq& \frac{r}{\sqrt{2}}-\frac{r^{3}}{4\sqrt{2}}+O(r^{5}), \nonumber \\ 
f_{K} &\simeq& \frac{r^{2}}{\sqrt{2}}-\frac{r^{6}}{4\sqrt{2}}+O(r^{10}), \\ \nonumber
f_{B}&\simeq& \sqrt{0.3437}r-0.2501\frac{r^{3}}{2}\sqrt{0.3437}+O(r^{5}); 
\end{eqnarray}
while the behavior for large $r$ is
\begin{eqnarray}
f_{F}&\simeq& 1-\frac{1}{r^{2}}+O\left(\frac{1}{r^{4}}\right), \nonumber \\ 
f_{K}&\simeq& 1-\frac{1}{r^{4}}+O\left(\frac{1}{r^{8}}\right),\\ \nonumber
f_{B}&\simeq& 1+\frac{2}{11r^{2}}+O\left(\frac{1}{r^{4}}\right). \label{large_eta_ber1}
\end{eqnarray}
Notice that, even though $f_{B}\rightarrow 1$ as $r \rightarrow \infty$,  the Berloff profile, $f_{B}$, exceeds unity somewhat in an intermediate range of $r$, which is qualitatively incorrect. 

By imposing the Fetter and Kerr profiles as initial conditions and by evolving them through the GP equation in a periodic domain it is also discovered that, as a consequence of being less accurate approximants, they are markedly unstable: see Fig. 2 of Meichle {\it et al.}\cite{Meichle}
Indeed, the core region first relaxes toward the exact profile while the excess energy released is successively propagated outwards as waves. These remain localized in the outer region of the domain, where they persist and are responsible for background oscillations which one might be tempted to identify as `thermal noise'. If these vortex profiles are imposed for two-vortex initial conditions, and vortex annihilation or vortex reconnection are studied, it is observed that the presence of these background `acoustic' waves strongly affects the vortex trajectories and the speed of the process.\cite{mythesis}  In particular, vortex trajectories become more wavy and annihilation or reconnection take longer to occur. One may speculate that in the presence of a `thermal background' some effective frictional effects are generated; we comment on this further in the following section.

Being a stationary solution of the GP equation, the exact single vortex profile minimizes the Hamiltonian. For this reason the value of the total energy is indicative of the  accuracy of the approximate profiles.
The energy per unit length of a single vortex with $n=1$ in a cylindrical domain of radius $R$ is given by \cite{Berloff}
\begin{eqnarray}
E'&=&\pi\left\{\int_{0}^{R}\left[\frac{df(r)}{dr}\right]^{2}r dr+\int_{0}^{R}\frac{f^{2}(r)}{r}dr\right. \nonumber \\
&&+\left. \frac{1}{2}\int_{0}^{R}[1-f(r)^{2}]^{2}rdr\right\},
\label{E_1v}
\end{eqnarray} 
where the first term gives the quantum energy $E'_{Q}$, the second, the classical kinetic energy $E'_{CK}$ and the third, the potential energy $E'_{I}$.

For large $R$ we have $f(r)\simeq 1$ and (\ref{E_1v}) simply  gives $E'\approx \pi \ln{R}$. For greater asymptotic precision Pitaevskii  \cite{pitaevskii} found $E'=\pi [\ln(R)+L_{0}]$, with $L_{0}=0.38$. The values of the parameter $L_{0}$ for all the five profiles and for the numerically diffused profile are reported in Table \ref{L0}. 

\section{Vortex dipole dynamics}
\begin{figure*}[h!tbp] \centering
\resizebox{16cm}{!}{\includegraphics{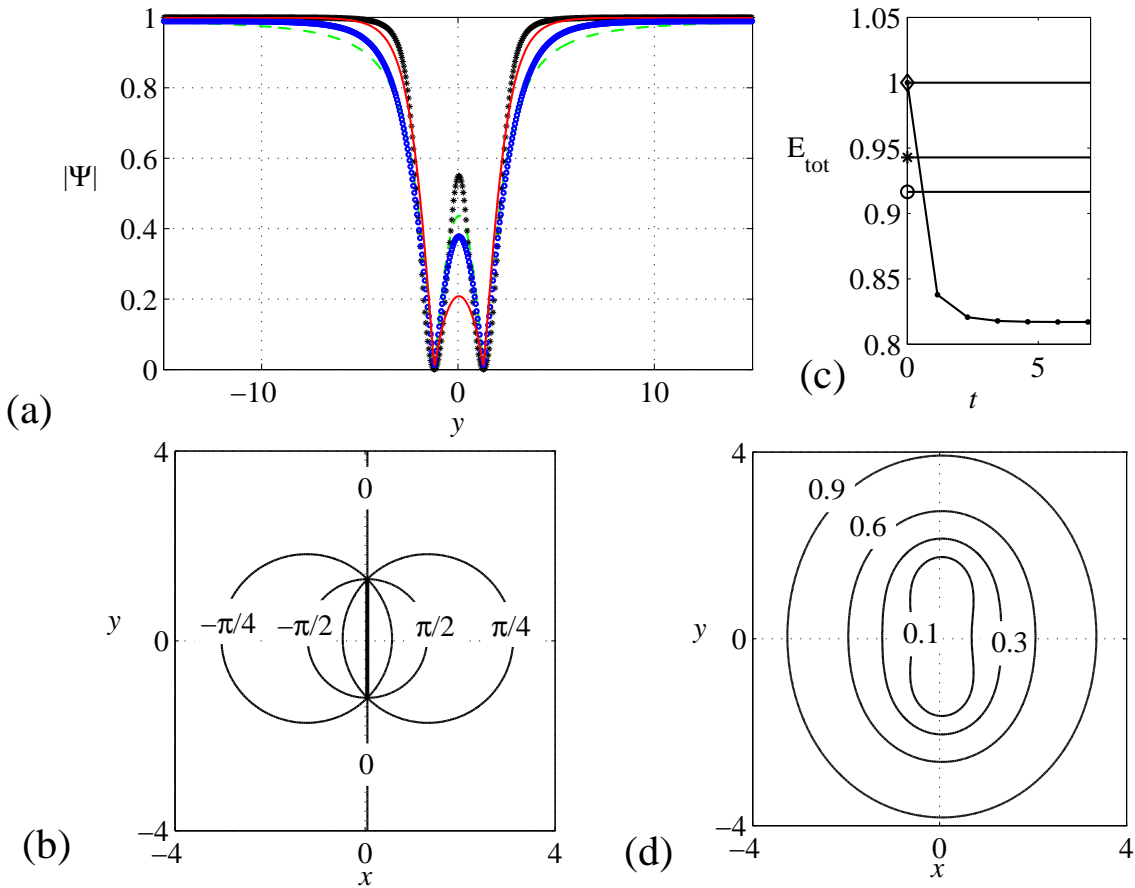}}
\caption{\label{comp_2vortices} (a) Two-vortex profiles viewed along ($x=0$, $y$), for $d_0=2.5$, obtained by multiplying wave-functions for each of the two vortices as in (\ref{molt}): the Fetter profile (green, dashed line), the Kerr profile (black stars) the Berloff profile (blue circles); or, by relaxing through the DGP equation any one of the three previous two-vortex profiles while keeping the initial phase contours fixed (red, solid line); (b) phase contours given by (\ref{phase}), associated to the profiles shown in (a) and plotted every $\pi/4$; (c) total energy per unit mass for the two-vortex Fetter profile (diamond), Kerr profile (star), Berloff profile (circle), and time decay of the total energy observed letting the DGP equation evolve from the two-vortex Fetter profile (dotted line). In this plot the maximum value is normalized to unity;  (d) density  distribution, i.e., $|\Psi|^{2}$ contours, for the diffused profile.}
\end{figure*}

\begin{figure*}[h!tbp] \centering
\resizebox{7cm}{!}{\includegraphics{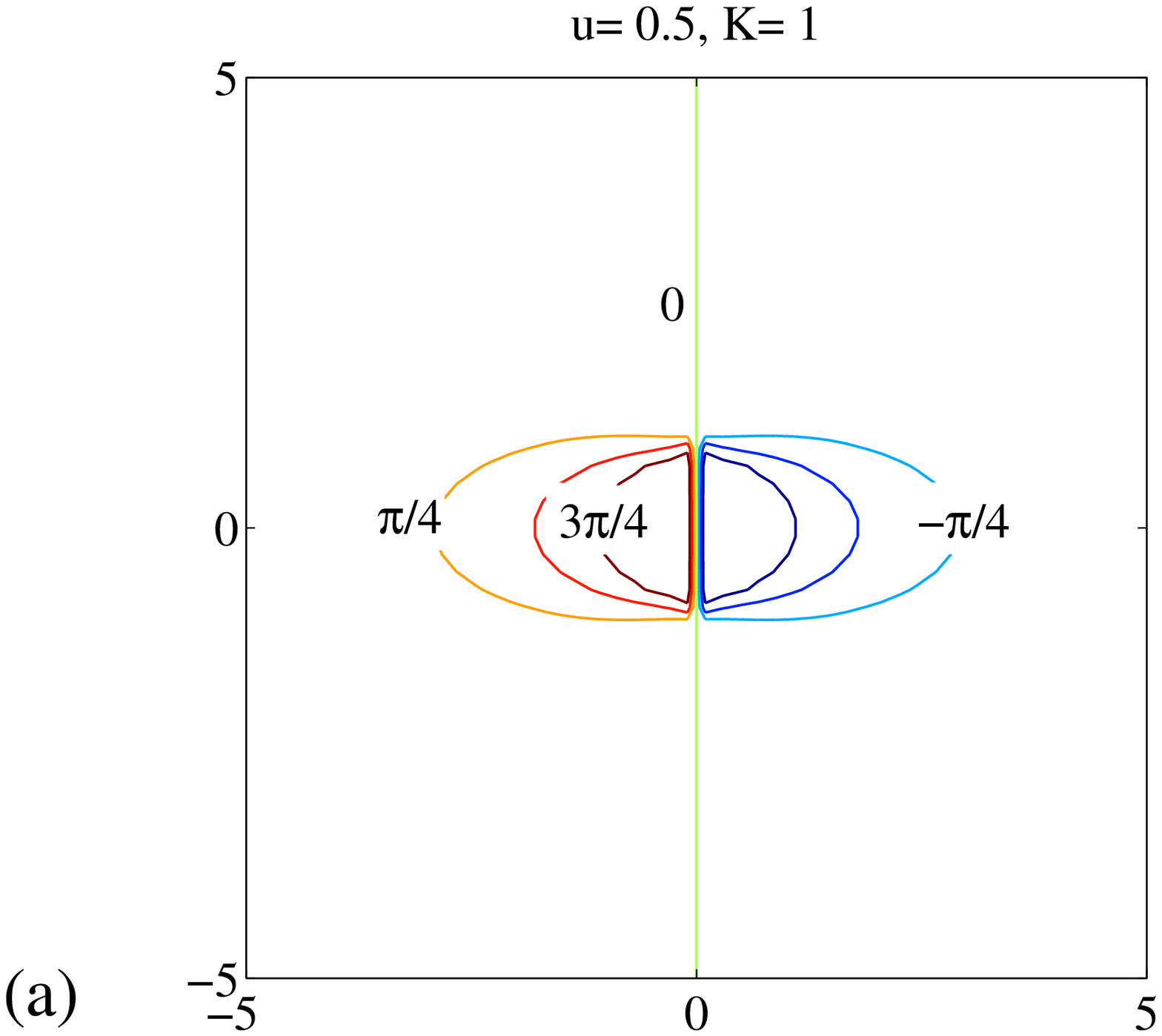}}
\resizebox{7cm}{!}{\includegraphics{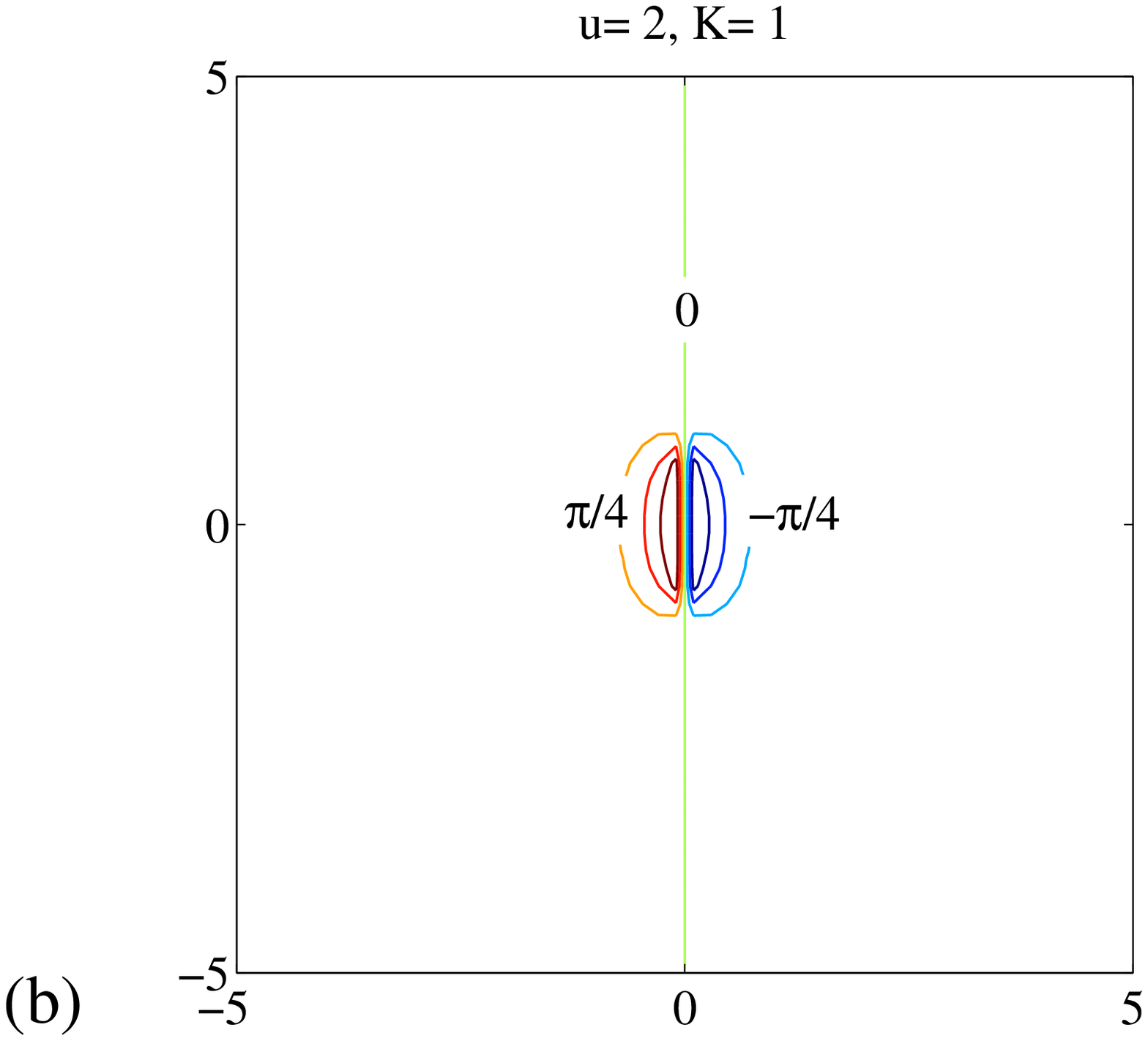}}
\resizebox{7cm}{!}{\includegraphics{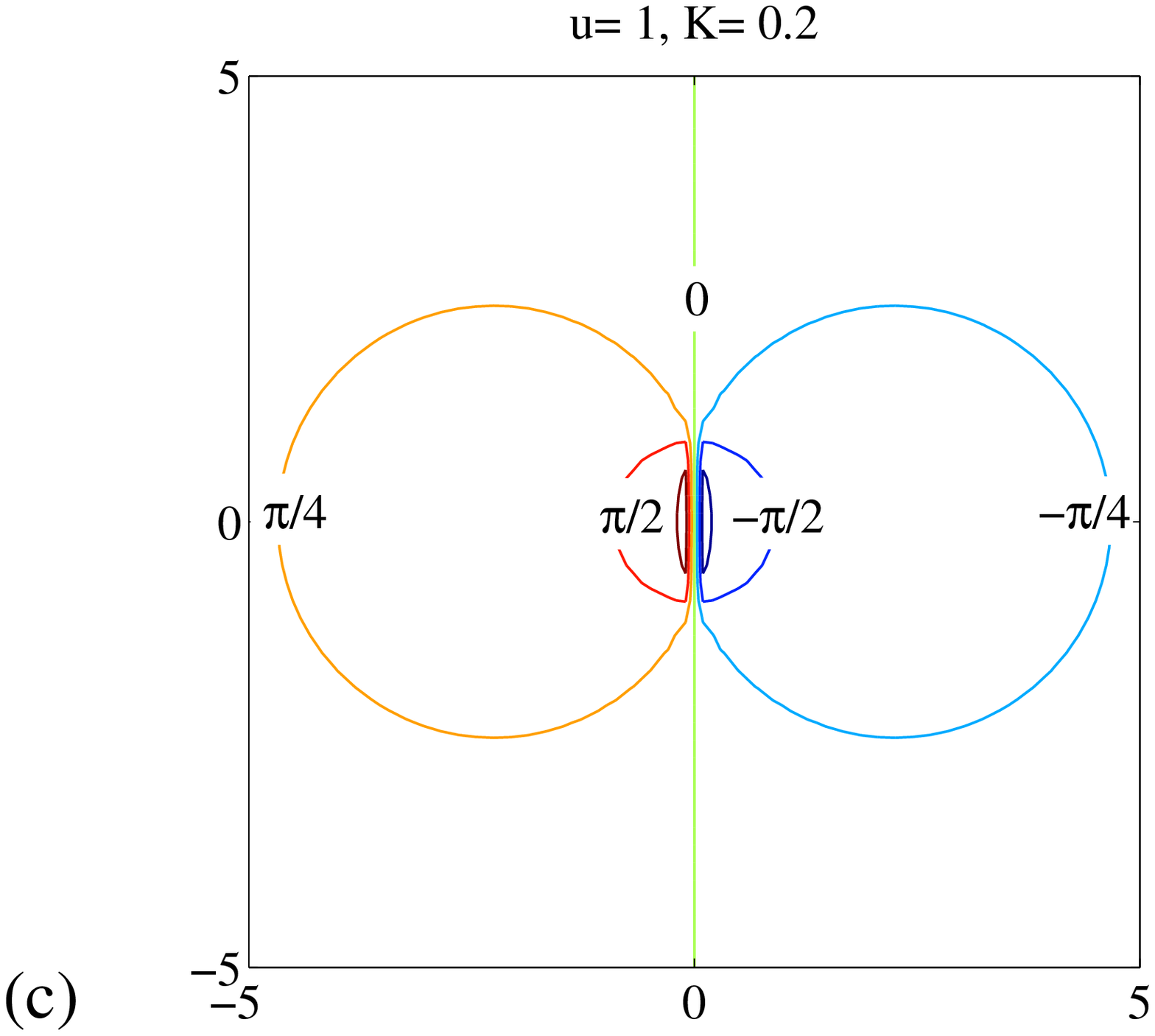}}
\resizebox{7cm}{!}{\includegraphics{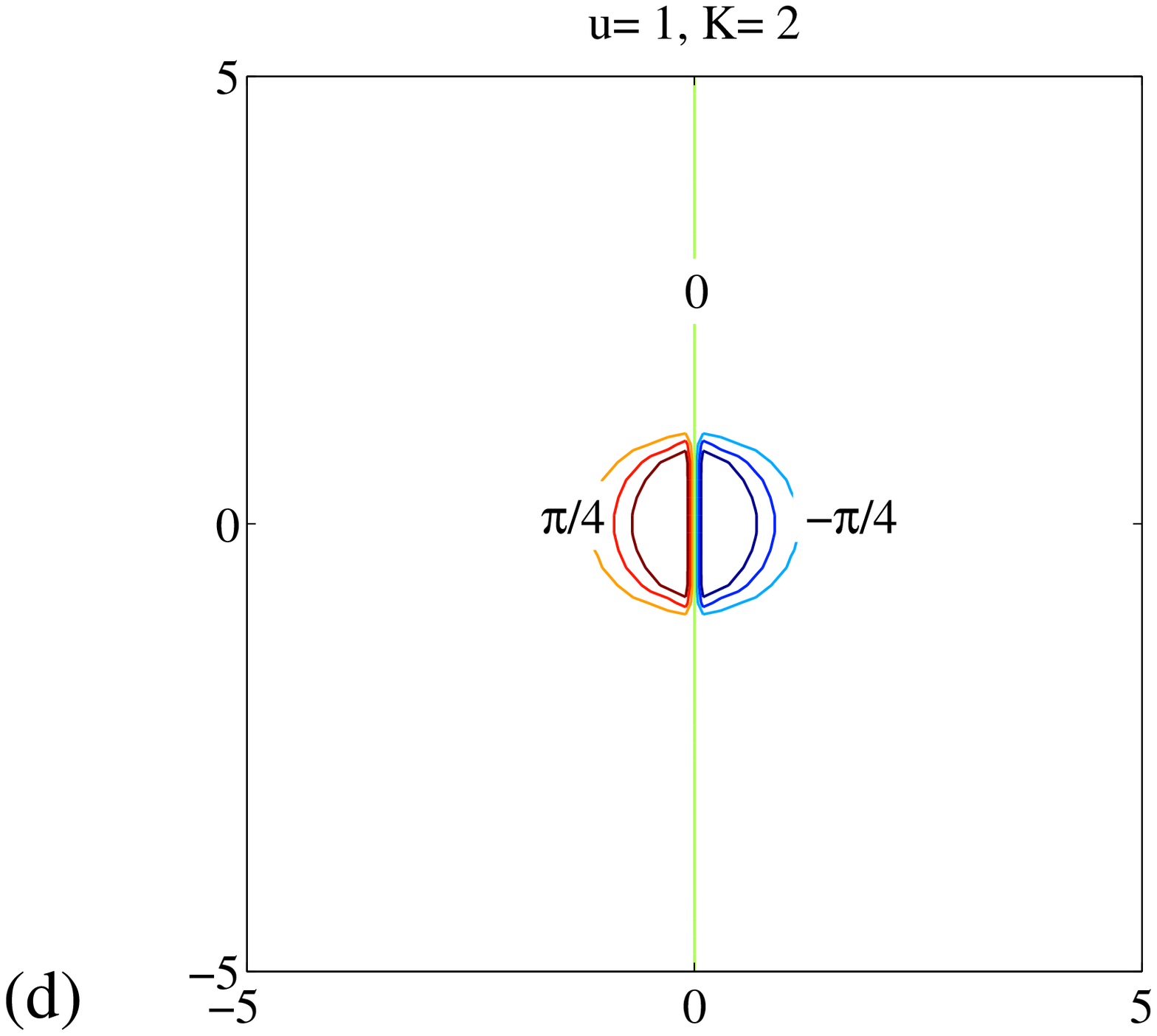}}
\caption{\label{elliptic_phases} Phase contours given by eq. (\ref{phase_ellipt}) and plotted every $\pi/4$ for different values of the elliptical parameters $u$ and $K$.}
\end{figure*}

Customarily, as mentioned in the Introduction, initial conditions for multiple vortices have been constructed simply by taking the product of the wave functions for each individual vortex,\cite{koplik93, koplik96,Tsubota02} that is 
\begin{equation}\Psi_{0}({\bf x},t=0)=\prod_{j}^{}f_{j}e^{i\phi_{j}},\label{molt}
\end{equation} 
where $f_{j}({\bf x})$ and $\phi_{j}({\bf x})$ are the profile and phase of the $j$-th vortex with specified core-locus. 
In the case of a vortex dipole characterized by a separation distance $d_0$, different two-vortex profiles provide different choices of $f_{j}$ [see Fig. \ref{comp_2vortices}(a)], while the sum of the phases $\phi_{j}$ leads for all cases to the `circular' phase contours, illustrated in Fig. \ref{comp_2vortices}(b). We have, in addition, taken account of the periodic images in accord to the explanations at the end of Sec. II above.\cite{mythesis}

In our vortex dipole calculations we have used a 2D domain of size $L_{x}=L_{y}=50$, spanning the coordinate range
\begin{equation}
-25 \geqslant x \geqslant 25 \mbox{        ,        } -25 \geqslant y \geqslant 25,
\end{equation} 
with $x=0$ and $|y|=a=d_{0}/2$ setting the location of the vortex cores.
The initial global phase $\phi_{0}$, according to (\ref{molt}), is then 
\begin{eqnarray}
\phi_{0}(x,y;t=0)&=&\phi_{1}+\phi_{2} \nonumber \\
&=&\tan^{-1}\left( \frac{y-a}{x}\right)-\tan^{-1}\left( \frac{y+a}{x}\right), \nonumber \\
&=&\tan^{-1}\left( \frac{-2ax}{x^2+y^2-a^2}\right).
 \label{phase}
\end{eqnarray}

Now consider an alternative approach in which: ({\it i}) the phase contours (\ref{phase}) are generalized to the elliptical distributions
\begin{equation}
\phi_{0}(x,y; t=0)=\tan^{-1}\left\{ \frac{-2aux}{K[(ux)^2+y^2-a^2]}\right\},
 \label{phase_ellipt}
\end{equation}
with $u$ and $K$ as parameters;
({\it ii}) the profiles for a trial initial condition are obtained using (\ref{molt}), but  are relaxed through the application of (\ref{diff}), namely the DGP equation [see Fig. \ref{comp_2vortices}, parts (c) and (d)], while the phase is {\it fixed} by (\ref{phase_ellipt}). Clearly the values $K=u=1$ correspond to the `circular' profile given by (\ref{phase}).

In (\ref{phase_ellipt}) the loci of fixed $\phi_0$ are ellipses that pass through the dipolar cores at ($x$, $y$)=(0, $\pm \frac{1}{2}d_{0}$) with axes parallel to the $x$ and $y$ axes. When $\phi_0=\pm \pi/2$ the ellipse is centered at the origin and has width $x_{0}=\pm (a/u)$. If $u \rightarrow 0$ this ellipse degenerates into a pair of  parallel horizontal straight lines. When $\phi_0$ is small the center of the ellipse is displaced to $x_0\approx-1/Ku\phi_0$ and the width increases by a factor $\sqrt{1+1/\phi_0 K u}$. Finally, for $u<1$ the ellipse is elongated along the $x$ axis, but for $u>1$ along the $y$ axis, while, when $K<1$ the phase contours at fixed $\phi_0$ are further apart than when $K=1$. See Fig. \ref{elliptic_phases} for an illustration.

Note that two antiparallel vortices forming a vortex dipole are not a stationary solution of the GP equation. Hence if such a configuration is allowed to diffuse fully through the DGP equation, it converges to the flat solution, $\bar{\Psi}=1$. However, when phase contours are enforced, i.e., held fixed while diffusing, a two-vortex combined profile relaxes to a lower energy solution which preserves the presence of both vortices as illustrated in Fig. \ref{comp_2vortices}. 

In the split-step spectral method each time increment $\Delta t$ is divided into two steps: the first integrates the nonlinear term in  the Cartesian frame of reference, the second integrates the linear term in spectral space. The desired phase $\phi_{0}$ is imposed by redefining, after each step, a wave function $\Psi_{\mbox{\footnotesize{enforced  phase}}}$ given by the modulus, $|\Psi|$, of the wave function just calculated times the exponential of $\phi_{0}$ as specified in (\ref{phase_ellipt}) so that  $\Psi_{\mbox{\footnotesize{enforced  phase}}}=|\Psi|e^{i\phi_{0}}$. 

A hint as to the outcome one may expect from the time evolution of such initial conditions is obtained by analyzing the linearized GP equation,\cite{Nazarenko} which, in fact, is simply the time-dependent one-particle Schr\"odinger (OPS) equation. By applying a phase shift $\Psi=\Phi e^{it/2}$ and neglecting the nonlinear term, the GP equation reduces to the OPS equation which, in terms of $\Phi({\bf x}, t)$, is
\begin{equation}
2i\frac{\partial \Phi}{\partial t}+\nabla^{2}\Phi=0. \label{OPS}
\end{equation}

Given the elliptical phases (\ref{phase_ellipt}) the chosen initial condition becomes
\begin{eqnarray}
\Psi_{0}&=&\frac{(ux)^2+y^2-a^2+2iaxu/K}{\mathcal{P}(r)}
            \label{circle_zeros}
\end{eqnarray} 
where $\mathcal{P}(r)$ is any appropriate polynomial that allows for the correct asymptotic behavior of $|\Psi_{0}|$ for $r$ close to and far from the vortex cores. For the sake of illustration and to simplify the calculations we take $\mathcal{P}=1$; in doing so, however,  the correct asymptotic behavior is not guaranteed, especially for $r$ far from the vortex cores since $|\Psi|\rightarrow \infty$ as $r \rightarrow \infty$.
In (\ref{circle_zeros}) the imaginary and real parts of $\Psi_{0}$ are simultaneously zero at ($x=0$,  $y=\pm a$), which, indeed, provides the location of the vortex lines. 

The solution of (\ref{OPS}) with $\Phi({\bf x}$, $t$=0$)=\Psi_{0}$, is found to have real and imaginary parts given by
\begin{equation}
\mbox{Re}(\Phi)=(ux)^2+y^2-a^2, \label{OPSE_circle_sol}
\end{equation}
$$\mbox{Im}(\Phi)=2axu/K+ t (u^2+1),$$
which simultaneously vanish when 
\begin{eqnarray}
x&=&- t(u^2+1)K/2au, \label{OPSE_circle}\\ \nonumber
y_{1,2}&=&\pm\sqrt{a^2-\left[\frac{t(u^2+1)K}{2a}\right]^2}. 
\end{eqnarray}
Hence, in this limit, the vortex cores move linearly with time in the direction of negative $x$ along the ellipse which coincides with Re$(\Psi_{0})=0$. On equating the first space-derivative of $|\Phi({\bf x}, t)|^2$ to zero one finds two solutions: the first reproduces (\ref{OPSE_circle}) and corresponds to the moving cores, i.e., the zero-density points; but the second one is given by $y=0$  and 
\begin{equation}
u^4x^3+a^2u^2x(2/K-1)+aut(u^2+1)/K=0,
\end{equation}
which can be solved using Cardano's formula for the roots of a cubic equation. The real root of this equation  corresponds to the location of the minimum density points {\it after annihilation} of the vortex dipole.

We conclude that, given the elliptical phase initial configuration, a vortex dipole with $\mathcal{P}$ constant or of first order in ($x$,$y$) should be expected to annihilate via the linearized GP equation in accord with (\ref{OPSE_circle_sol}). If $\mathcal{P}$ is of higher order, (\ref{OPSE_circle_sol}) represents an approximate solution accurate to first-order in time. The approach that lead us to find solution (\ref{OPSE_circle_sol}) has been proposed and followed by Nazarenko and West.\cite{Nazarenko}

Observe that if $u \rightarrow 0$ and $uw=K \rightarrow 0$, that is, $u$ and $K$ go to zero, but in fixed proportion with $w$ as the control parameter, the condition (\ref{circle_zeros}) becomes
\begin{equation}
\Psi_{0}=(y^2-a^2)+2iaxw. \label{straight_zeros}
\end{equation}
The OPS solution for this is 
\begin{equation}
\Phi({\bf x}, t)=y^2-a^2+i(t+2axw).
\end{equation}
The time behavior of the vortex core is then given by 
 \begin{equation}
y_{1,2}=\pm a \:\:\: \mbox{     and      } \:\:\: x= -t/2aw.
\end{equation}  
Thus the vortex dipole now does {\it not} annihilate but, rather, moves with a fixed velocity of propagation inversely proportional to $a$ and $w$.

In the following we solve numerically the nonlinear GP equation for different vortex-dipole initial conditions and we show:  ({\it a}) how annihilation of a vortex dipole may be affected by the choice of different initial profiles $f_{j}$; ({\it b}) how, given `circular' phase contours and a diffused two-vortex profiles, the solution depends on the initial separation distance $d_{0}$; and ({\it c}) how, the vortex-dipole dynamics is influenced by varying the initial phase configurations via the elliptical parameters $u$ and $K$ in (\ref{phase_ellipt}).  
\subsection{Sensitivity to choice of initial profiles}
 \begin{figure}[h!tbp] \centering
\resizebox{8.5cm}{!}{\includegraphics{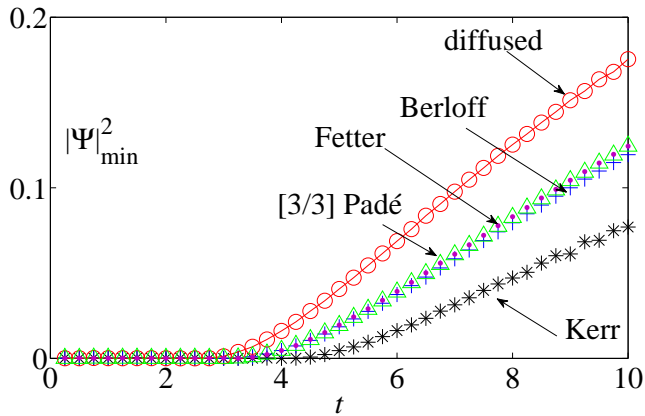}}
\caption{\label{d3_min_dens} Time evolution of the minimum density for two antiparallel vortices with initial separation $d_{0}=2.5$, starting from four different initial conditions: the two-vortex Fetter (violet dots), Kerr (black stars), Berloff (blue crosses), and [3/3] Pad\'{e} (green triangles) profiles and any of the previous conditions diffused through the DGP equation, (\ref{diff}), keeping the phase contours fixed (red open circles).}
\end{figure}

\begin{figure}[h!tbp] \centering
\resizebox{8cm}{!}{\includegraphics{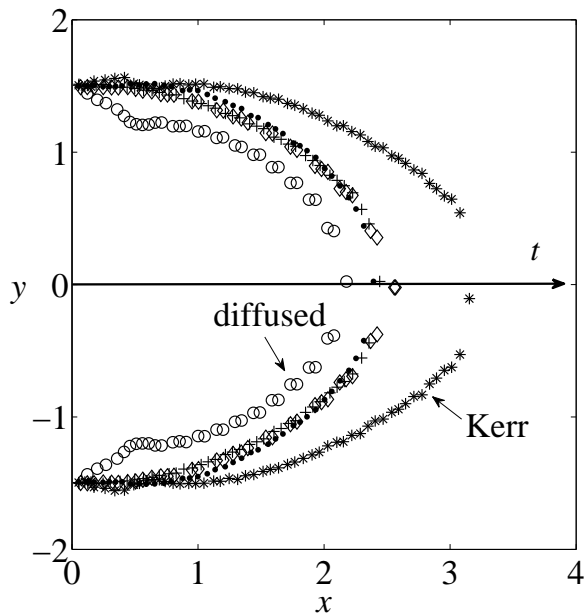}}
\caption{\label{x_y_trajectories} Position in the ($x$, $y$) plane of the vortex cores as time proceeds, comparing the cases when the initial profiles are derived from the Kerr (stars), Berloff (crosses), Fetter (dots), [3/3] Pad\'{e} approximants (diamonds), with the relaxed or diffused two-vortex profile (circles).}
\end{figure}

\begin{figure*}[h!tbp] \centering
\resizebox{17cm}{!}{\includegraphics{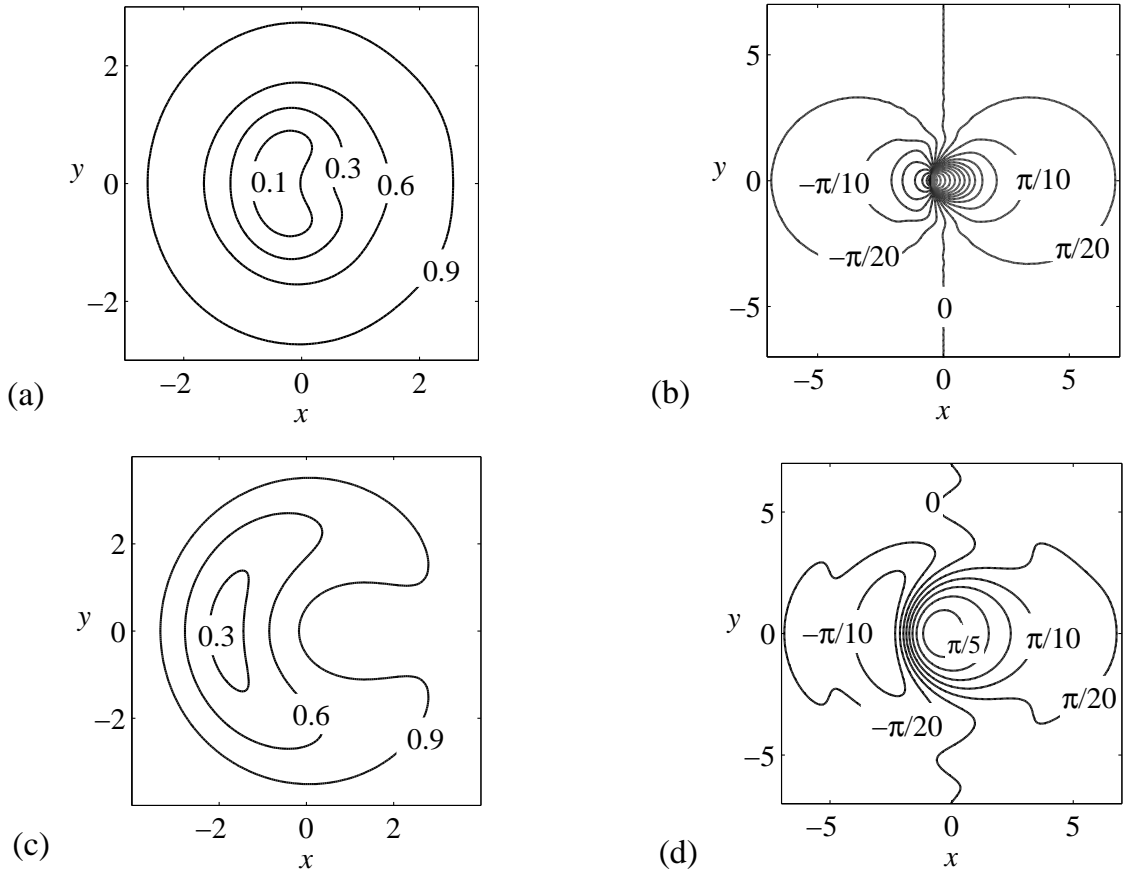}}
\caption{\label{4_images} Annihilation process through the GP equation for $d_{0}=1$ showing only a small portion of the domain. At $t=0.25$: (a) distribution of $|\Psi|^2$; (b) phase contours at multiples of $\pi/20$. At $t=2$: (c) distribution of $|\Psi|^2$ after annihilation, the minimum density exceeding zero;  (d) phase distribution now confined to the interval $-\pi/20\lesssim \phi \lesssim pi/5$.}
\end{figure*}

\begin{figure*}[h!tbp] \centering
\resizebox{12cm}{!}{\includegraphics{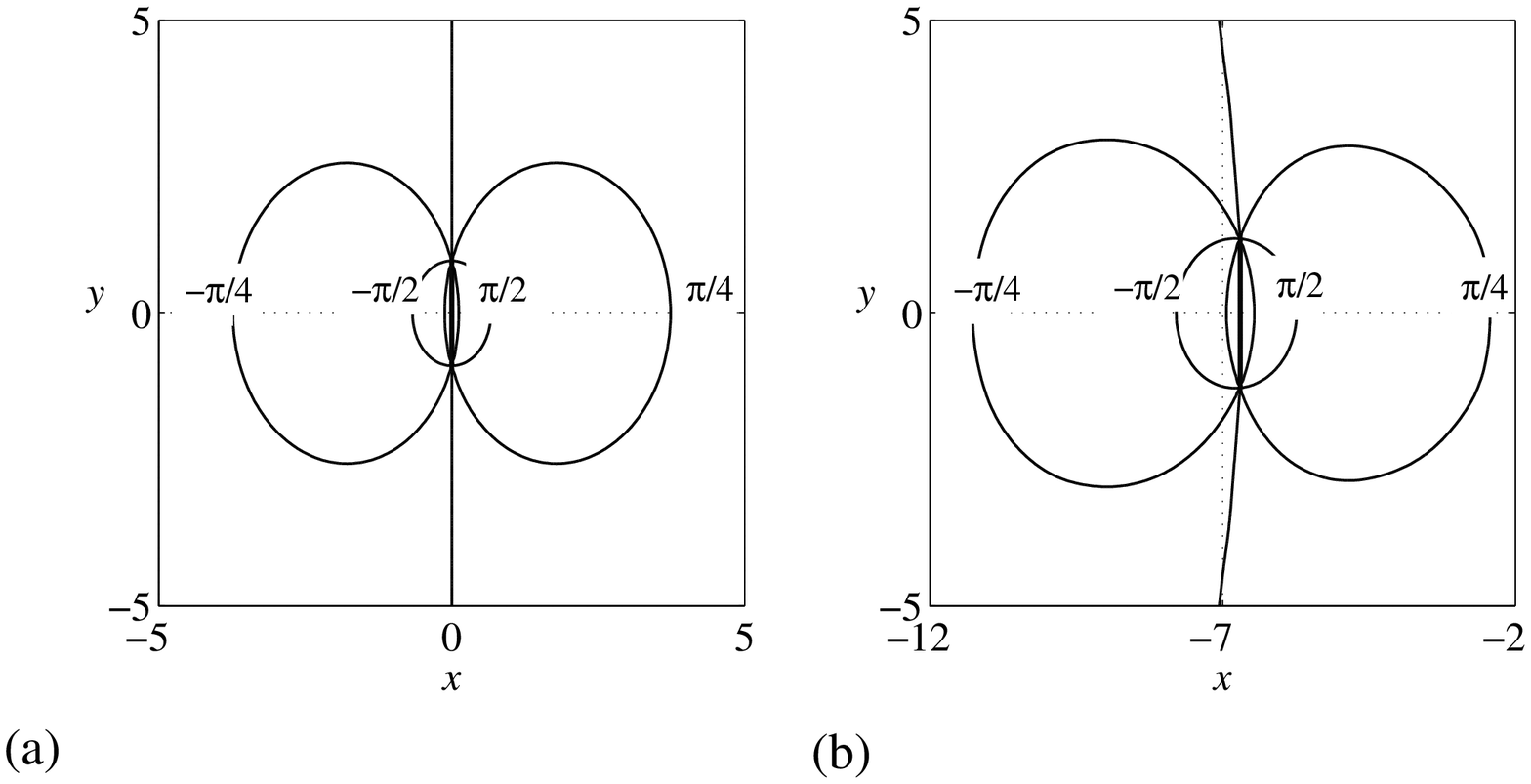}}
\caption{\label{phi_stationary_wave} (a) Phase distribution of the solitary-wave approximant derived by Berloff \cite{berloff2} for steady velocity $U$=0.4, to be compared with (b) the phase distribution at $t$=20 for an initial condition characterized by $d_0=4$ after evolution by the GP equation.}
\end{figure*}

In the first case we take $K=u=1$ in (\ref{phase_ellipt}), i.e., circular phase contours, and examine different two-vortex initial profiles. We find that if the diffused structure (at fixed phase distribution) for a vortex dipole is imposed as an initial condition for the GP equation, the annihilation mechanism proceeds at a markedly different rate than when the initial condition is obtained by multiplying the two single-vortex profiles: see the open-circles plot in Fig. \ref{d3_min_dens}. Evidently the diffused profile commences annihilating sooner and the annihilation process proceeds more rapidly than do the other three profiles. This may be rationalized by noting that the non-diffused profiles possess excess energy that needs to be radiated away from the cores before the annihilation itself can get properly underway. 

In confirmation of this  interpretation, the Kerr profile is the one which starts the annihilation process most slowly, followed in order by the Fetter, Berloff, [3/3] Pad\'{e}, and diffused profiles.  
The detailed trajectories of the vortex centers before annihilation are also strongly influenced by the choice of the initial condition as illustrated in Fig. \ref{x_y_trajectories} for the case $d_{0}=3$. For these simulations the conservation of the total energy $\Delta E$ is accurate to within 0.3\%.

The dipole annihilation mechanism can be thought to proceed  generally in two stages: ({\it i}) the two vortices translate and move towards one another while the phase reduces its range of variation everywhere except close to the vortices: see Fig. \ref{4_images}(a) and (b); ({\it ii}) the cores merge together,   annihilation occurs and the minimum density increases while the whole structure continues to move forward radiating acoustic waves: Fig. \ref{4_images}(c) and (d). This description is consistent with the results reported by Ogawa {\it et al.}\cite{Tsubota02}

By choosing a two-vortex diffused profile and repeating the calculations for different initial separations, it is found that annihilation always occurs when $d_{0} \lesssim 3$. If, instead, the initial separation exceeds $d_{0} \simeq 4$ the vortex dipole eventually propagates as a solitary wave (albeit usually with associated oscillations). It transpires, however, that this observation is deeply connected with the specific choice of the initial phase distribution.
We note, indeed, that the phase contours entailed in the approximate solitary wave solutions found by Berloff \cite{berloff2} for $d_0=1.8$ are more-or-less elliptical in shape, rather than circular, as those imposed here:  see Fig. \ref{phi_stationary_wave}(a). Interestingly, when our $d_0=4$ initial condition is evolved through the GP equation, it behaves as a solitary wave and its phase distribution, which is initially circular as in Fig. \ref{comp_2vortices}(c), relaxes to an elliptical-like configuration rather similar to Berloff's result as shown in Fig.  \ref{phi_stationary_wave}(b).

\subsection{Dependence on the initial separation distance $d_{0}$}
\begin{figure*}[h!tbp] \centering
\resizebox{16cm}{!}{\includegraphics{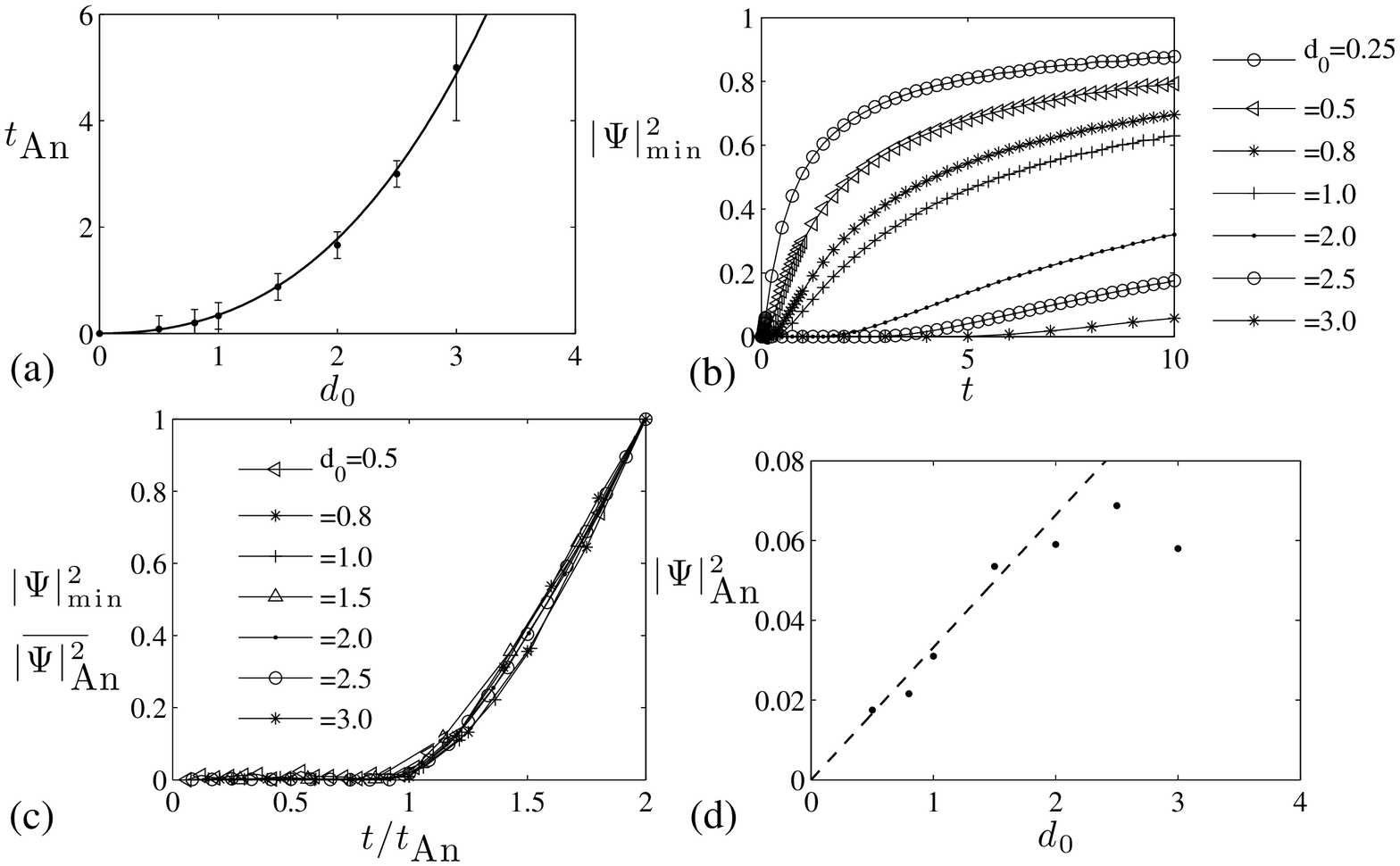}}
\caption{\label{psi2_t} Vortex dipole GP annihilation dynamics for circular phase contours ($K=u=1$) relaxed via the DGP equation.
(a) The annihilation time $t_{\mbox{\footnotesize{An}}}(d_0)$, and a simple fit, quadratic as $d_0\rightarrow 0$: for details see text. (b) The minimum density $|\Psi|_{\min}^2$ as a function of time as the vortex dipoles merge; (c) the collapse of the $|\Psi|_{\min}^2$ plots rescaled by $t_{\mbox{\footnotesize{An}}}(d_0)$ and by the characteristic density, $|\Psi|_{\mbox{\footnotesize{An}}}^2$, shown in (d) as a function of $d_0$. See further in the text.}
\end{figure*}

Starting from circular phase contours ($K=u=1$) and a two-vortex diffused profile we first report in Fig. \ref{psi2_t}(a) the the {\it annihilation time}, $t_{\mbox{\footnotesize An}}(d_{0})$, needed to just complete the merging of the two distinct cores. The approximately quadratic dependence of $t_{\mbox{\footnotesize An}}$ on $d_{0}$ observed might have been anticipated from the solution (\ref{OPSE_circle}) of the linearized GP (or OPS) equation (\ref{OPS}). Indeed, this yields an annihilation time
\begin{equation}
t_{\mbox{\footnotesize{An}}}^{0}(d_{0})=b_{0}d_{0}^2\:\:\:\:\: \mbox{with}\:\:\:\:\:b_0=1/2K(u^2+1), \label{lin_theory}
\end{equation}
where the superscript denotes the linearized derivation, despite which the result seems likely to be generally valid for $d_0 \rightarrow 0$.
Accordingly in Fig. \ref{psi2_t}(a), where $K=u=1$, we present a fit using
\begin{equation}
t_{\mbox{\footnotesize{An}}}(d_0)\simeq \frac{1}{4}d_0^2[1+b_1d_0]
\end{equation}
with $b_1=0.39$. Clearly these limited data are consistent with the form (\ref{lin_theory});  however, while recognizing the relatively large uncertainties involved in these calculations,  it should, perhaps, be mentioned that a direct estimate of  $b_0$  yields a somewhat larger value around 1/3. 

Once the vortex dipole has been annihilated there remains only a `dip' characterized by a minimal density, $|\Psi(t)|^2_{\mbox{\footnotesize{min}}}$, which, as $t$ grows above $t_{\mbox{\footnotesize{An}}}$, increases monotonically from zero but initially fairly slowly: see the plots in Fig. \ref{psi2_t}(b), especially those for $d_0>1.5$. Then, for larger times, the density minimum evidently approaches a plateau. The process depends strongly on the initial separation $d_0$ in  a manner that seems linked to the total missing mass associated with original vortices, namely, $\int_{A}(1-|\Psi|^2) dA$. Furthermore, the visually quite varied plots in Fig. \ref{psi2_t}(b) can be reduced to a single `master growth curve', as seen in Fig. \ref{psi2_t}(c), by rescaling the time by $t_{\mbox{\footnotesize{An}}}(d_0)$ and the density by a characteristic density $|\Psi(d_0)|_{\mbox{\footnotesize{An}}}^{2}$. To determine $|\Psi|_{\mbox{\footnotesize{An}}}^2$ we used the value $|\Psi(d_0; t=2t_{\mbox{\footnotesize{An}}})|^2$ which, as reported in Fig. \ref{psi2_t}(d), initially rises linearly with $d_0$ with a slope (indicated by a dashed line) of about 0.033.

\begin{figure*}[h!tbp] \centering
\resizebox{8.5cm}{!}{\includegraphics{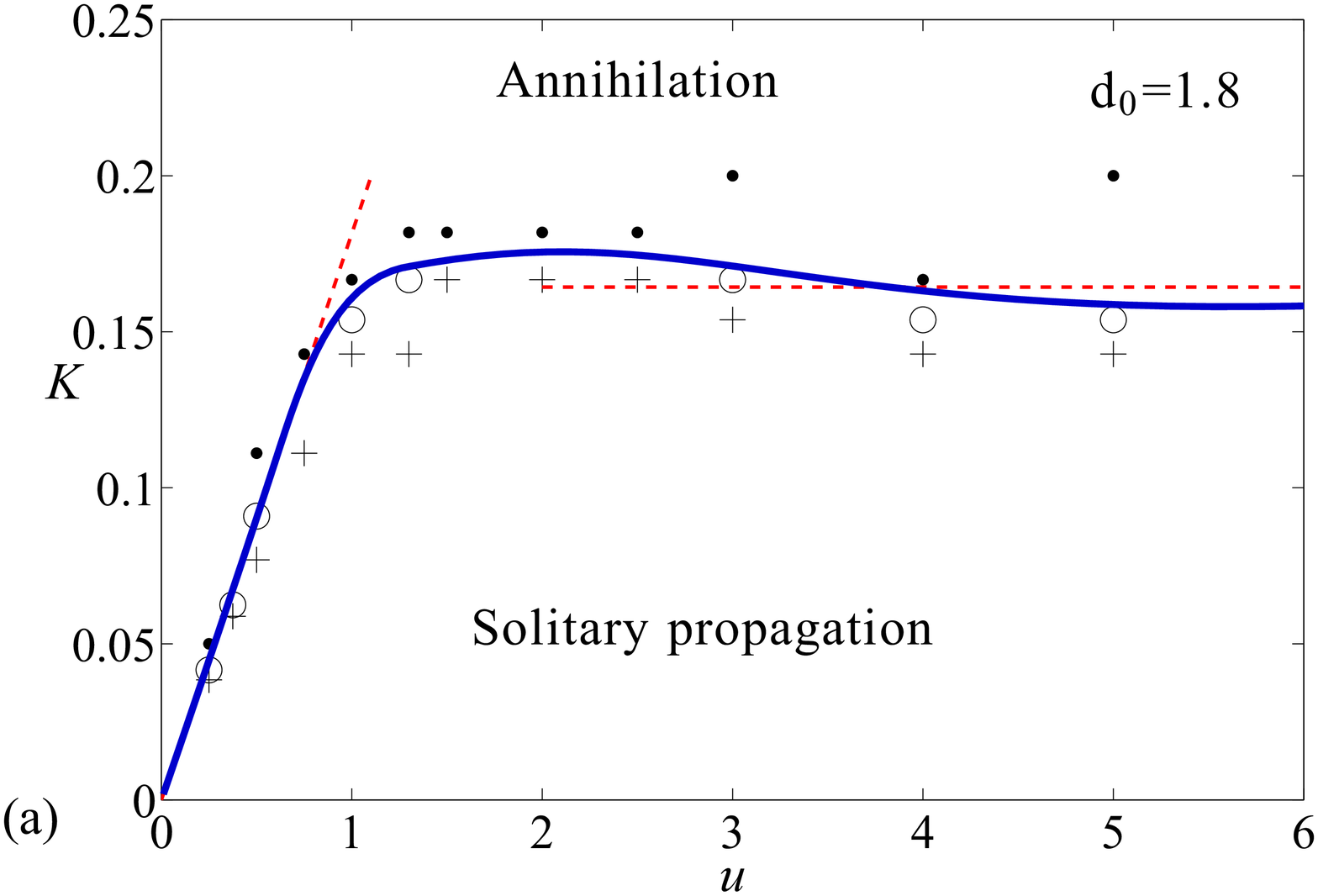}}
\resizebox{8.5cm}{!}{\includegraphics{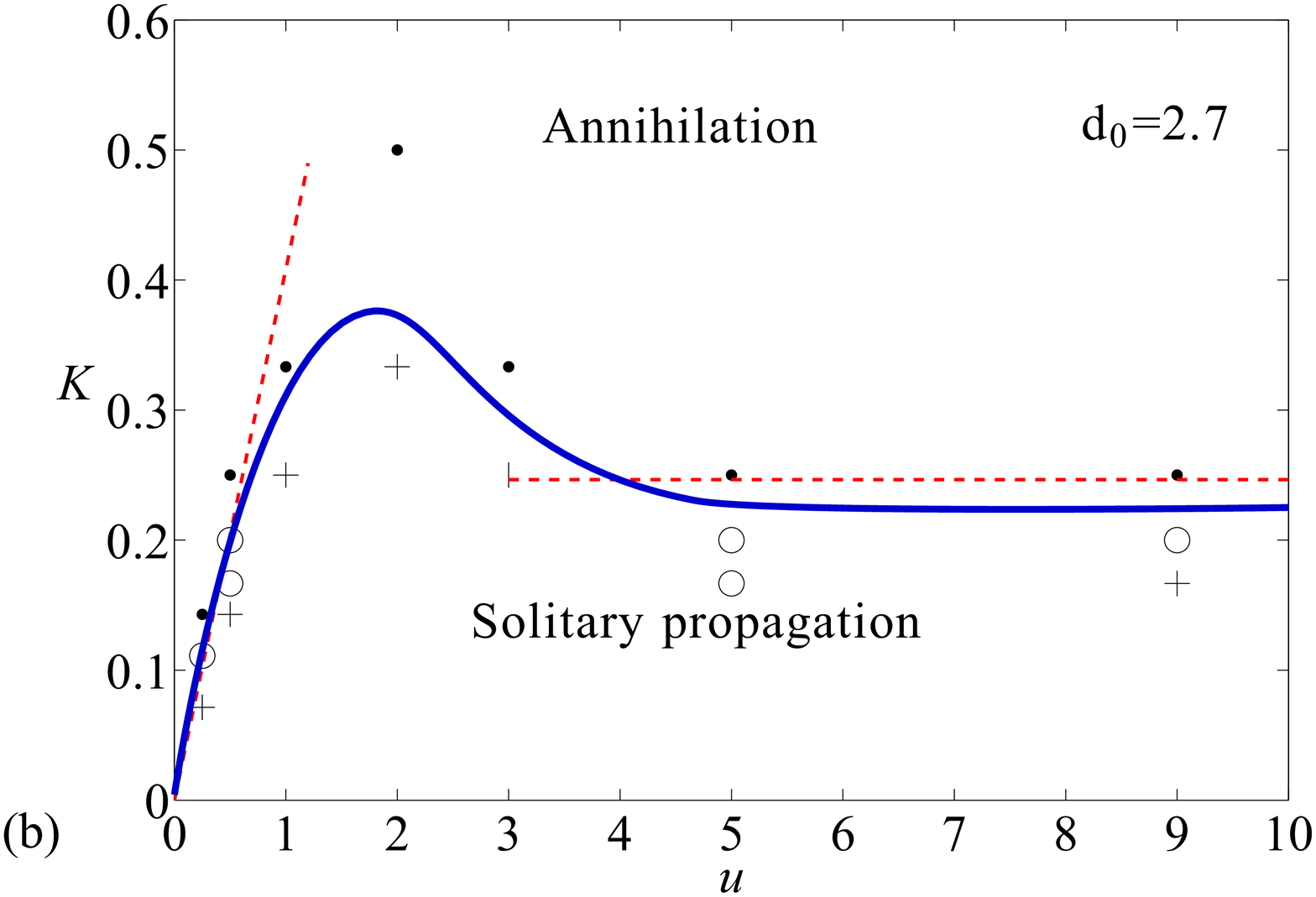}}
\resizebox{9cm}{!}{\includegraphics{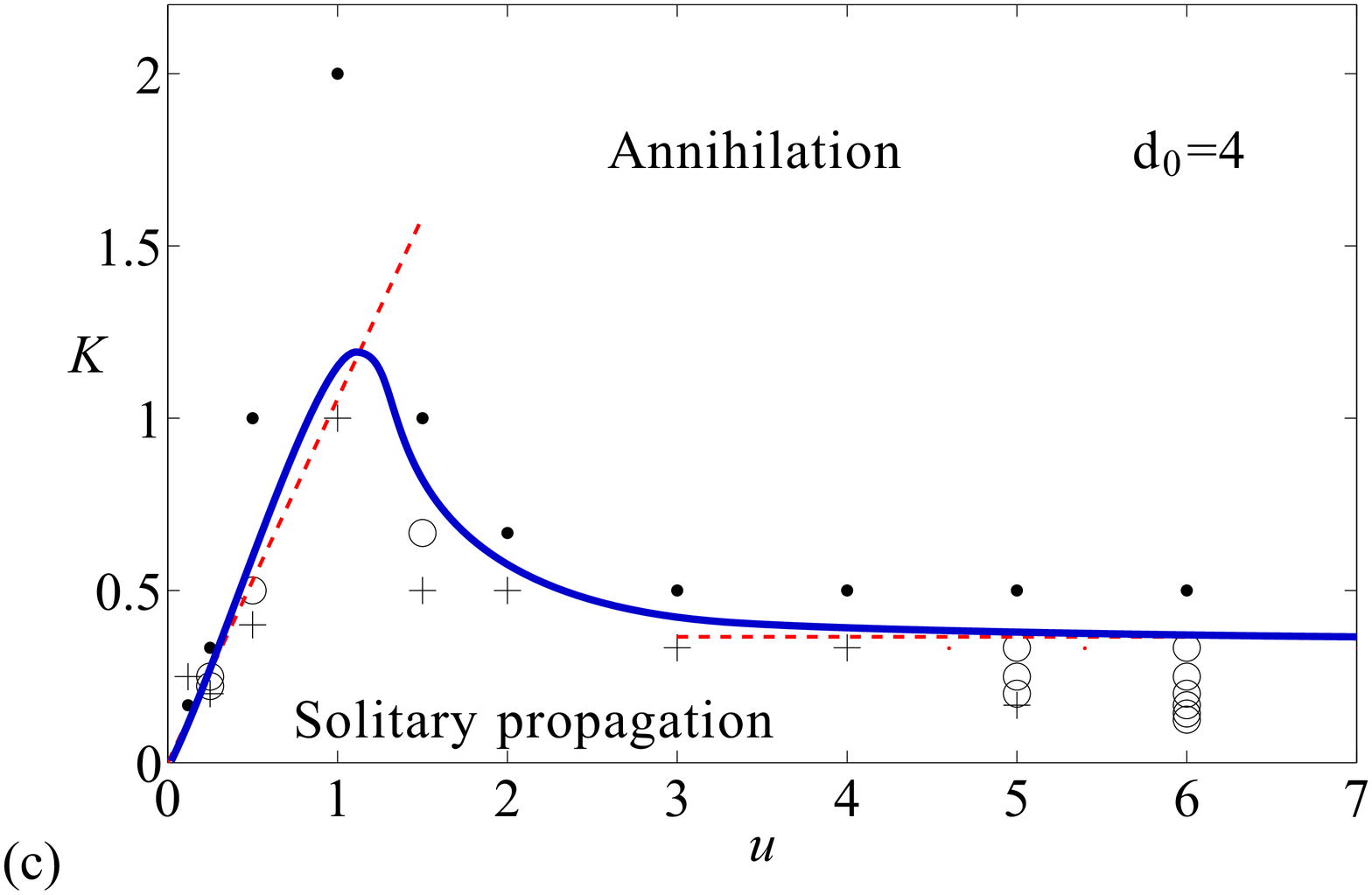}}

\caption{\label{ann_prop} Annihilating and propagating behavior for a vortex dipole characterized by elliptical phase contours specified by ($K$, $u$) for separations: (a) $d_0=1.8$, (b) $d_{0}=2.7$, and (c) $d_{0}=4.0$. Dots ($\bullet$) correspond to annihilating dipoles, and crosses (+) to propagating dipoles. The circles (o) indicate calculations for parameters close to the AnSol boundary in which annihilation is almost observed, as the vortex cores approach closely, although, subsequently, a strongly oscillating propagating dipole appears to develop. The solid lines (color online) are guides to the eye as regards the AnSol boundaries.}
\end{figure*}
\subsection{Sensitivity to initial phase configuration}

As our last study a set of initial conditions, with profiles relaxed via the DGP equation, are imposed by enforcing the elliptical phase contours stated in (\ref{phase_ellipt}) for a range of the parameters $u$ and $K$. The resulting transition between the annihilating and propagating behavior of the vortex dipoles is evident in Fig. \ref{ann_prop} for the three separation values: (a) $d_0=1.8$, which is the separation for which Berloff derived a solitary-wave approximant;\cite{berloff2} (b) $d_0=2.7$ and (c) $d_0=4.0$, the value for which we originally observed solitary-wave behavior with initially circular phase contours. The annihilation-solitary wave (or AnSol) boundary clearly depends on both the ellipticity parameters, $K$ and $u$, and on the separation $d_0$ between the parallel vortices.

Note, first, that the data suggest quite strongly that the AnSol boundary approaches the ($K$, $u$) origin {\it linearly} when $u$ (and $K$) become small. In this limit we find the boundary is reasonably well represented by
\begin{equation}
K_c(u; d_0)\simeq c_0 u d_0^2,
\end{equation}
with $c_0\simeq 0.056$ for $d_0\lesssim4$. On the other hand, a theoretical justification of this expression even, say, for $d_0 \rightarrow 0$ (for example on the basis of solutions of the linearized GP equation or the corresponding solitary wave GP equation\cite{berloff2}) appears nontrivial.

Then on noting the distinct vertical scales in Fig. \ref{ann_prop} we see that for large $u$ the AnSol boundary becomes independent of $u$ so that $K_c(u; d_0)\rightarrow K_{\infty}(d_0)$ as $u \rightarrow \infty$. However the larger the vortex separation $d_0$, the larger the value of $K$ at which the AnSol transition is realized. In fact it seems that
\begin{equation}
K_{\infty}(d_0)\simeq c_{\infty}d_0,
\end{equation}
with $c_{\infty}\simeq 0.091$ provides, within rather large uncertainties, a tolerable description.

Lastly, it appears that at least for $d_0\gtrsim 2$, the AnSol boundary is nonmonotonic in $u$; rather $K_c(u; d_0)$ displays a maximum around $u=$ 1-2 which, relative to $K_{\infty}(d_0)$, also increases quite dramatically with $d_0$. More extensive calculations would be needed, however, to provide more quantitative conclusions.

\section{Conclusions}
A single straight vortex is usually regarded as a stationary axisymmetric solution of the Gross-Pitaevskii equation. We have demonstrated, however,  that some approximate vortex profiles proposed in the literature contain a significant excess of energy which, under time development, leads to the outwards propagation of compressional waves.  These effects prove quite small for the particular Pad\'{e} approximant proposed by Berloff \cite{berloff2} and are certainly negligible for our newly proposed exact two-point [2/2] and [3/3] Pad\'{e} approximants.\cite{Baker} Thus these profiles qualify as good candidates to perform low-energy calculations: a desirable condition, since the background `thermalization' which arises from less accurate profiles is an uncontrolled and, initially at least, an unwanted effect. Furthermore, the undetermined excess energy interferes with the precise quantitative measurements one should hope to perform. It may be expected, moreover,  that real physical systems in which energy dissipation (not included in the GP model) is to be anticipated, will tend to assume minimal energy configurations when feasible. 

The second part of our study has been devoted to the dynamics of vortex dipoles, characterized by a separation distance $d_{0}$ between two parallel counter-rotating infinite straight line vortices. It was demonstrated that such a dipole can {\it either} annihilate {\it or} propagate steadily as a solitary wave. Both the qualitative and the quantitative behavior depend strongly on the {\it initial vortex profile} and on the {\it initial phase contours}. Starting from an initial condition obtained through multiplication of two single-vortex wave functions one typically observes annihilation for an initial separation, $d_0$, smaller than three intrinsic units, but solitary-wave behavior for an initial separation larger than four units. When $d_0$ is small, a good approximation for the dynamical behavior is given by the linearized GP equation. The linear term then dominates and annihilation is observed, as predicted by the one-particle Schr\"odinger equation. If lower energy initial profiles are used, annihilation occurs faster. The $d_{0}$ dependence of the solutions has been explored by determining the low-energy initial conditions via the Diffuse GP equation and the time evolution via the GP equation.

When the separation distance is larger, the linearized solution is no longer valid, the nonlinear term plays a role and its  interplay with the linear terms determines the solitary wave behavior; of course, this is a typical nonlinear phenomenon as observed, e.g., in the Korteweg-de Vries equation.\cite{hirota, ablowitz} It is interesting to note that, in this case, starting with a rough, approximate solitary-wave as the initial condition, the solution rapidly descends, as time develops, into the `solitary-wave valley' in an idealized landscape map of possible solutions. On the other hand, when the initial separation distance is small, this outcome is rarely observed. However, solitary wave solutions for small distances, even for the limiting case $d_{0}=0$, do exist; but they require precise initial conditions such as those found by Berloff.\cite{berloff2} 

More quantitatively, we have thus studied a class of explicit initial phase configurations that yield elliptical phase contours specified by two parameters, $K$ and $u$: see the expression (\ref{phase_ellipt}) and Fig. \ref{ann_prop} which reveals well-defined annihilation vs. solitary wave (or AnSol) boundaries of characteristic form.

In conclusion, we remark, in more general terms, that vortices are specified by the intersections of the zeroes of the real and imaginary parts of the effective wave function: these form `zero lines' (or linear loci) in two spatial dimensions, or `zero planes' (or planar manifolds) in three dimensions. There are many possible topologies, and functional forms associated with them, that allow for vortex {\it intersections}. Fixed points of the GP equation then correspond to well defined topologies; \cite{Meichle} but for dynamical structures such as a vortex dipole or reconnecting vortices the choice of initial configurations remains arbitrary and may strongly influence the outcome. One must conclude that the custom of representing the various topologies merely by the product of approximate wave functions derived from straight vortices is likely to be a significant limitation in developing a valid perspective for understanding the entire dynamic landscape more fully.
\section*{Acknowledgements}
We thank Daniel P. Lathrop, Robert M. Kerr, David P. Meichle, and Enrico Fonda for their scientific interest and useful discussions.
Cecilia Rorai gratefully acknowledges support from Universit\'{a} degli Studi di Trieste, University of Maryland and New York University. 
\bibliography{arxiv_rorai_sreeni_fisher.bbl}
\end{document}